\documentclass[aps,%
reprint,
superscriptaddress,
amsmath,amssymb,floatfix,
prl
]{revtex4-2}
\usepackage{lipsum}
\usepackage{graphicx}
\usepackage{dcolumn}
\usepackage{comment}
\usepackage{bm}
\usepackage{color}
\usepackage{graphicx}
\usepackage{dcolumn}
\usepackage{comment}
\usepackage{bm}
\usepackage{xcolor}

\definecolor{darkblue}{rgb}{0,0,0.6}
\usepackage{hyperref}
\hypersetup{colorlinks,linkcolor=darkblue,citecolor=darkblue,urlcolor=darkblue}

\usepackage[normalem]{ulem}

\begin{document}

\title{Conservation laws and slow dynamics determine the universality class of interfaces in active matter}

\author{Raphaël Maire}
\email{maire@ub.edu}
\affiliation{Universit\'e Paris-Saclay, CNRS, Laboratoire de Physique des Solides, 91405 Orsay, France}
\author{Andrea Plati}
\affiliation{Universit\'e Paris-Saclay, CNRS, Laboratoire de Physique des Solides, 91405 Orsay, France}
\author{ Frank Smallenburg}
\affiliation{Universit\'e Paris-Saclay, CNRS, Laboratoire de Physique des Solides, 91405 Orsay, France}
\author{Giuseppe Foffi}
\email{giuseppe.foffi@universite-paris-saclay.fr}
\affiliation{Universit\'e Paris-Saclay, CNRS, Laboratoire de Physique des Solides, 91405 Orsay, France}

\date{\today}

\begin{abstract}
While equilibrium interfaces display universal large-scale statistics, interfaces in phase-separated active and driven systems are predicted to belong to distinct non-equilibrium universality classes. Yet, such behavior has proven difficult to observe, with most systems exhibiting equilibrium-like fluctuations despite their strongly non-equilibrium microscopic dynamics. We introduce a hard-disk model driven by active collisions, conceived as an effective 2D description of a vibrofluidized granular system that, contrary to self-propelled models, displays clear non-equilibrium interfacial scaling. We observe for the first time, the $|\boldsymbol q|$KPZ and wet-$|\boldsymbol q|$KPZ universality classes while revealing a new, previously overlooked universality class arising in systems with slow solid-like or glassy dynamics. Conservation laws and slow dynamics select these distinct classes.
\end{abstract}
\maketitle

\emph{Introduction -} At thermodynamic equilibrium, the Boltzmann distribution governs all static observables~\cite{kardar2007statistical}. For example, every liquid–gas critical point---whether in a colloidal suspension, a molecular fluid, or the $\mu VT$ ensemble---shares the same static critical exponents~\cite{goldenfeld2018lectures} while the dynamics are captured by different dynamical exponents $z$~\cite{mazenko2006nonequilibrium,tauber2014critical}. Similarly, interfaces in equilibrium phase-separated systems---such as those found in liquid–liquid colloid demixing~\cite{aarts2004direct}, polymer blends~\cite{langevin2021light}, electrolyte solutions~\cite{safouane2009surface}, or 2D Ising models~\cite{gallavotti1972phase}---are all governed by the capillary-wave Hamiltonian on large scales with surface tension $\gamma$: $\mathcal{H}[h] \simeq \frac{\gamma}{2} \int d^d x \left( \bm{\nabla} h \right)^2$, where $h(\bm{x}, t)$ is the height of a $d$-dimensional interface~\cite{onuki2002phase}.  Dynamically, projection methods show that the Fourier components $h(\bm{k}, t)$ evolve as~\cite{onuki2002phase,kawasaki1982kinetic, slavchov2021characterization, bray2001interface, bausch1981critical, zia1985normal, zia1988dynamics, Romano_Golestanian_Mahault_2025, caballero2020bulk}:
\begin{equation}
\partial_t h(\bm{k}, t) = -\kappa |\bm{k}|^{\alpha-2}\dfrac{\delta \mathcal H}{\delta h} + \sqrt{2\kappa |\bm{k}|^{\alpha-2}D} \zeta(\bm{k}, t),
\label{eq: dynamics equilibrium}
\end{equation}
where $D$ is proportional to the temperature, $\zeta(\bm{k}, t)$ is a unit-variance Gaussian white noise, and $\kappa |\bm k|^{\alpha-2}$ is a damping. The dynamical exponent $z$ characterizing the interface relaxation time equals $\alpha$, which depends on the conservation laws of the underlying $(d+1)$-dimensional order parameter dynamics~\cite{bray2001interface}. The fluctuation dissipation theorem ensures that static properties quantified by the static height correlation $S_h(\bm k)\equiv \langle |h(\bm k)|^2\rangle\sim \bm k^{-d-2\chi}$ with $\chi = (2-d)/2$, are independent of $z$ at equilibrium~\cite{onuki2002phase}. This decoupling between statics and dynamics is, however, only guaranteed at equilibrium.

With the rise of interest in active matter~\cite{te2025metareview, marchetti2013hydrodynamics, cates2024active} and the ubiquity of interfaces in biology~\cite{foty2005differential, hyman2014liquid, kammeraat2025correlated, schamberger2023curvature, sabari2020biomolecular, brangwynne2009germline, molliex2015phase}, research on far-from equilibrium interfaces has surged. In this work, we focus not on growing interfaces, such as those described by KPZ~\cite{kardar1986dynamic, takeuchi2010universal}, which make up a large class of physical systems, but rather on non-growing interfaces in phase-separated systems. Freed from a Hamiltonian framework, these interfaces can exhibit peculiar behavior such as negative surface tension~\cite{bialke2015negative,fausti2021capillary, singh2019hydrodynamically, cates2024active}, instabilities~\cite{zhao2024asymmetric,alert2022fingering, onuki2002phase, adkins2022dynamics, martinez2025interfacial}, and traveling waves~\cite{gulati2024traveling, barthelet1998benjamin, langford2024phase, rasshofer2025capillary}. These effects can typically be captured either by introducing effective parameters into the equilibrium interface equation (Eq. \eqref{eq: dynamics equilibrium}) or by coupling this equation to another field. Surprisingly, the measured fluctuations in granular or scalar active matter interfaces often follow the equilibrium result: $S\sim |\bm k|^{-2}$~\cite{luu2013capillarylike, yao2025interfacial, Wysocki_2016, langford2024theory, patch2018curvature, lee2017interface, paliwal2017non, del2019interface,sun2025interfacial, patteson2018propagation, coelho2020propagation, chacon2022intrinsic}. This equilibrium-like behavior remains puzzling. While microscopic violations of detailed balance do not always survive coarse-graining in other systems~\cite{lorenzana2025nonreciprocity, papanikolaou2024dynamic, caballero2020stealth, o2022time, plati2024self, crisanti2012nonequilibrium, fodor2022irreversibility,lucente2022inference}, theoretical predictions for interfaces in active matter suggest a wide variety of non-equilibrium behaviors depending on the conserved quantity of the underlying dynamics. For example, in overdamped systems, such as active Brownian particles, the interface is predicted to obey the $|\bm q|$KPZ equation~\cite{besse2023interface, toner2023roughening}:
\begin{equation}
    \partial_t h = -\kappa\gamma |\bm{k}|^{3}h+\lambda|\bm k|\mathcal F\left[(\bm \nabla h)^2\right] + \sqrt{2\kappa |\bm{k}|D} \zeta,
    \label{eq: qkpz}
\end{equation}
where $\mathcal F$ denotes the Fourier transform and $\lambda$ controls the non-linearity strength. Despite extensive research in interfacial fluctuations of active matter interfaces~\cite{luu2013capillarylike, yao2025interfacial, Wysocki_2016, langford2024theory, patch2018curvature, lee2017interface, paliwal2017non, del2019interface, chacon2022intrinsic}, $|\bm q|$KPZ has been reported only once in numerical simulations of strongly confined active particles~\cite{krishan2025finite}. 
Including momentum conservation should yield the wet-$|\bm q|$KPZ equation~\cite{caballero2024interface}:
\begin{equation}
    \partial_t h = -\kappa\gamma |\bm{k}|h+\lambda|\bm k|^{-1}\mathcal F\left[(\bm \nabla h)^2\right] + \sqrt{2\kappa |\bm{k}|^{-1}D} \zeta,
    \label{eq: wet qkpz}
\end{equation}
which remains unobserved. Finally, for some non-equilibrium hyperuniform (HU) systems where the noise is momentum-conserving, but the deterministic evolution includes viscous drag so that momentum is not a locally conserved quantity~\cite{maire2025hyperuniformity, hexner2017noise, lei2024non}, the interface obeys:
\begin{equation}
    \partial_t h = -\kappa\gamma |\bm{k}|^{3}h+ \sqrt{2\kappa |\bm{k}|^2D} \zeta.
    \label{eq: laplacian}
\end{equation}
To date, only this last equation has been clearly observed numerically in large-scale systems~\cite{maire2025hyperuniform}.\\
For 1D interfaces, each of these cases has different exponents:
\begin{equation*}
\begingroup
\setlength{\arraycolsep}{6pt}
\renewcommand{\arraystretch}{1.5}
\begin{array}{@{}l@{\quad}l|@{\quad}c@{\quad}c@{\quad}c@{}}
\text{1D interface} & & \chi & z & \text{Ref}\\
\hline
\text{Equilibrium} & \text{Eq.}~\eqref{eq: dynamics equilibrium} & 1/2 & \alpha & \onlinecite{onuki2002phase}\\
|\bm q|\text{KPZ} & \text{Eq.}~\eqref{eq: qkpz} & 0.3-0.4 & 2.2-2.8 & 
\text{\onlinecite{besse2023interface}, \onlinecite{caballero2024interface}}\\
\text{wet-}|\bm q|\text{KPZ} & \text{Eq.}~\eqref{eq: wet qkpz} & 1 & 1 & 
\onlinecite{caballero2024interface}\\
\text{HU interface} & \text{Eq.}~\eqref{eq: laplacian} & 0 & 3 & 
\onlinecite{maire2025hyperuniform}\\
\end{array}
\endgroup
\end{equation*}
Thus, three predicted non-equilibrium phase-separating models with distinct dynamics yield different static roughness exponents, breaking equilibrium universality.

The contribution of this work is twofold. (i) We introduce a hard-disk model driven by active collisions, inspired by a vibrated granular system which, in contrast to regular self-propelled active matter, exhibits clearly and for the first time all three non-equilibrium universality classes theoretically predicted for interfaces separating coexisting liquid and gas phases. (ii) We show that the emergence of solid-like ordering or glassiness in one of the phases qualitatively modifies interfacial fluctuations, leading to new, previously overlooked universality classes. Taken together, these results reveal how conservation laws and slow relaxation processes can fundamentally reshape the static properties of interfaces in active matter systems.

\emph{Model -} Taking inspiration from a granular system~\cite{maire2024non, maire2025dynamical}, we model $N$ 2D hard disks of mass $m$ and diameter $\sigma$ in a periodic $L_x\times L_y$ box at packing fraction $\phi = N\pi\sigma^2/(4L_xL_y)$. Each particle $i$ follows the Langevin equation:
\begin{equation}
m\dfrac{d^2\bm r_i}{dt^2}=-m\Gamma \dfrac{d\bm r_i}{dt} + \sum_{j\neq i}\bm F_{ij}^{\rm coll}+\sqrt{2m\Gamma T}\bm \zeta_i(t),
\label{eq: time evol}
\end{equation}
where $\Gamma$ is a damping, $T$ is a kinetic temperature, and $\bm F_{ij}^{\rm coll}$ acts as a momentum-conserving non-equilibrium collision such that at contact ($|\bm r_i - \bm r_j| = \sigma$), the kinetic energy of $i$ and $j$ changes by~\cite{risso2018effective, maire2025hyperuniform}:
\begin{equation}
\Delta E_{ij} = \Delta E_{ij}^+-m\dfrac{1-\alpha^2}{4}(\bm v_{ij}\cdot \hat{\bm\sigma}_{ij})^2,
\label{eq: energy change at collision}
\end{equation}
with $\bm v_{ij}=\bm v_{i}-\bm v_{j}$, $\hat{\bm\sigma}_{ij}=(\bm r_i - \bm r_j)/|\bm r_i - \bm r_j|$, and $0\leq\alpha \leq1$ a coefficient of restitution. If $\Gamma = 0$, we set $\alpha < 1$ to dissipate the energy injected by $\Delta E_{ij}^+>0$. The injected energy depends on the times since each particle’s last collision $\tau_i$ and $\tau_j$:
\begin{equation}
\Delta E^+_{ij} = 2\delta E_0 + \delta E\left(1 - e^{-\tau_i/\tau_r}\right)^\beta + \delta E\left(1 - e^{-\tau_j/\tau_r}\right)^\beta,
\label{eq: energy recharging}
\end{equation}
where $2\delta E_0$ and $2(\delta E_0+\delta E)$ are respectively, the minimum and maximum possible injected energies. The terms $\big(1-e^{-\tau/\tau_r}\big)^\beta$ interpolate between these limits: $\tau_r$ sets the characteristic recharging time, while $\beta$ controls how sharp the approach to the maximum is. For large $\delta E$, dense regions cool and compress while dilute regions heat and expand, destabilizing the homogeneous phase and driving phase separation with an interface separating the two phases~\cite{risso2018effective, maire2025hyperuniform}. Physically, Eqs.~\eqref{eq: energy change at collision} and \eqref{eq: energy recharging} provide an effective 2D description of a vibrofluidized granular monolayer: vertical energy injected by a vertically vibrating plate is intermittently converted into horizontal motion during collisions~\cite{brito2013hydrodynamic}. The eliminated out-of-plane dynamics is encoded as a particle-level internal ``recharging'' state that controls $\Delta E^+_{ij}$. In this coarse-grained 2D picture, the activity emerges through collision-mediated energy transfer, similar to models with non-reciprocal interactions or chiral forces, where detailed balance is broken by interactions~\cite{Caprini2025bubble, Ouazan2023}.

We perform hybrid time-stepped/event-driven molecular dynamics simulations~\cite{smallenburg2022efficient} of Eqs.~\eqref{eq: time evol} and \eqref{eq: energy change at collision} in an elongated box ($L_x>L_y\equiv L$). Time and energy are measured in arbitrary units $\hat\tau$ and $\hat E=m(\sigma/\hat\tau)^2/2$, respectively. The system is initialized with a step density profile along $x$ at the coexistence densities of the two phases, and the interface position $h(y, t)$ is tracked over time (see Supplemental Material~\cite{supp}).

\begin{figure*}[t]
    \centering
\includegraphics[width=0.99\linewidth]{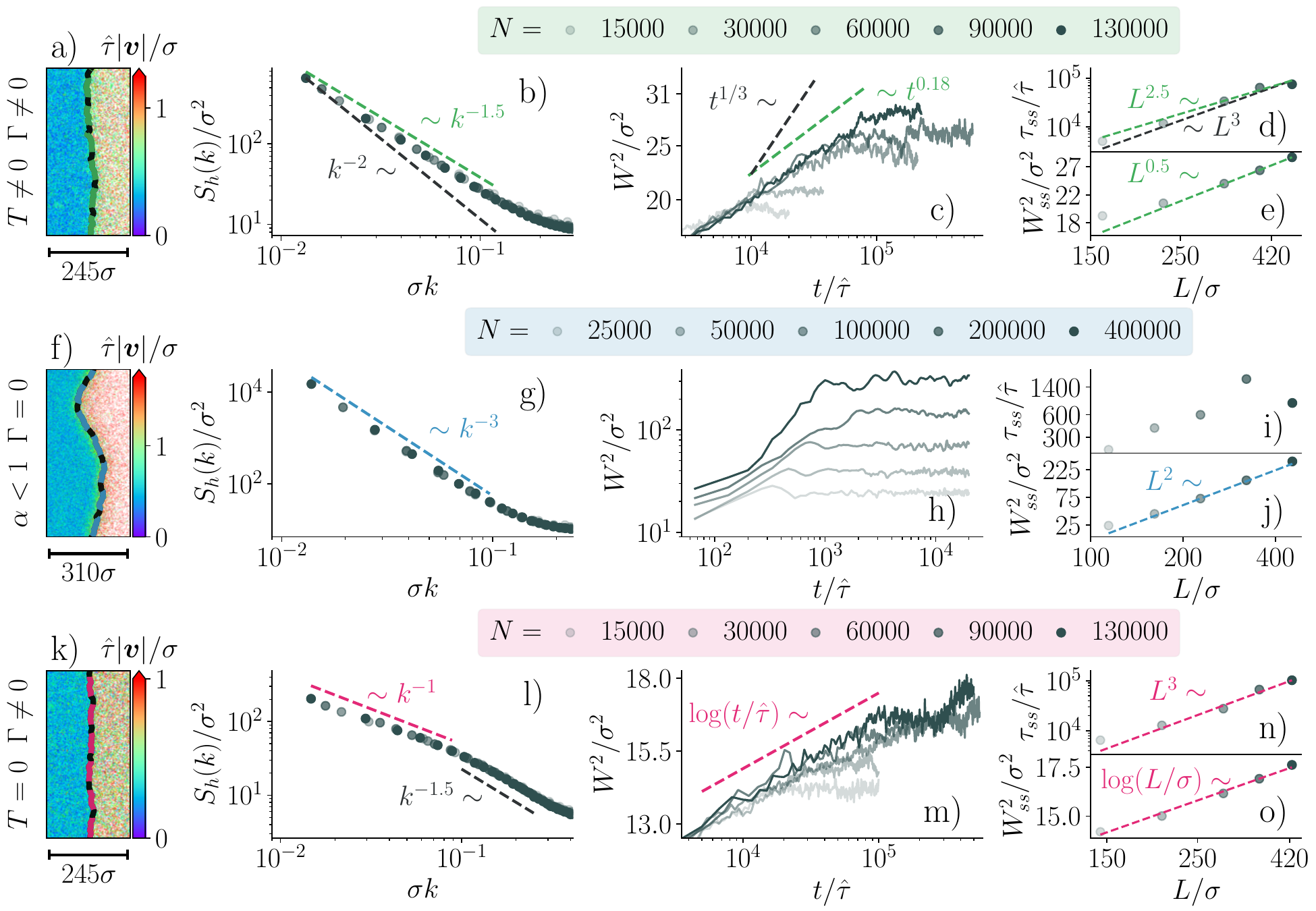}
    \caption{Simulations of three qualitatively distinct systems based on Eqs.~\eqref{eq: time evol} and \eqref{eq: energy change at collision}. In the first row, only the density field is conserved; in the second row, momentum is also conserved, and in the last row, only the density field is conserved, but the mesoscopic noise conserves the center-of-mass of the system. a), f), and k) show snapshots centered on the interface, with particles colored by velocity. Panels b), g), and l) display the steady-state static height correlations as a function of wavenumber. c), h), and m) show the time evolution of the interface width squared, starting from a flat interface, for various system sizes. d), e), i), j), n), and o) show the system-size dependence of the coarsening time and steady-state width squared $W^2_{ss}\equiv W^2(t\to\infty)$. The averages are computed over 50 to 1000 independent initial configurations, with at least 300 snapshots taken for each configuration. In a-e), we set $T/\hat E=0.05$, $\Gamma/\hat\tau=0.05$, $\tau_r/\hat\tau=4$, $\delta E_0/\hat E = 0.0025$, $\delta E/\hat E = 15$, $\beta = 10$, $\alpha = 0.99$ and  $L_x=2L_y$. For f-j), we set $\Gamma/\hat\tau=0$, $\tau_r/\hat\tau=3$, $\delta E_0/\hat E = 0.0125$, $\delta E/\hat E = 25$, $\beta = 10$, $\alpha = 0.95$ and $L_x=8L_y$. For k-o), we set $T/\hat E=0$, $\Gamma/\hat\tau=0.075$, $\tau_r/\hat\tau=3$, $\delta E_0/\hat E = 0.0125$, $\delta E/\hat E = 25$, $\beta = 10$ and  $\alpha = 0.95$, $L_x=2L_y$.}
    \label{fig: recap}
\end{figure*}

\emph{Numerical results - } The critical exponents are obtained from correlation functions starting from the flat interface of initial size $L$ at $t=0$~\cite{caballero2018strong}:
\begin{equation}
        W^2(L, t)\sim \left\{
        \begin{array}{ll}
            L^{2\chi} &  \mbox{if}\quad t\gg\tau_{ss}(L) \\
            t^{2\chi/z} & \mbox{if}\quad t\ll\tau_{ss}(L)
        \end{array},\quad\tau_{ss}(L)\sim L^z
        \right.
    \label{eq: observable}
\end{equation}
where $W^2\equiv\left\langle(h(y, t)-\langle h(y, t)\rangle_L)^2\right\rangle_L$ is the mean-squared width of the interface and $\tau_{ss}$ is the time required to reach stationarity. In practice, we estimate $\tau_{ss}$ from the intersection between the early-time fit $W^2\sim t^{2\chi/z}$ and the saturation value $W^2(L,t\to\infty)$~\cite{supp}. We will also use the static height correlation in the steady state $S_h(k)\equiv \langle|h(k, t\gg \tau_{ss})|^2\rangle_L\sim k^{-1- 2\chi}$ in 1D. The simulation results are summarized in Fig.~\ref{fig: recap}, with three limiting cases of our model shown in different rows. The average $\langle\cdot\rangle_L$ is over initial conditions with system size and flat interface length $L_y=L$, and, when applicable, multiple uncorrelated snapshots of the same run.

\textit{(i) $|\bm q|$KPZ scaling}. In the top row ($T\neq 0$ and $\Gamma \neq 0$), the Langevin bath damps the momentum, leaving the density as the only conserved field. This should place the resulting interface in the $|\bm q|$KPZ universality class where the following critical exponents are expected: $2.2 \lesssim z^{|\bm q|\rm KPZ} \lesssim 2.8$ and $0.3 \lesssim \chi^{|\bm q|\rm KPZ} \lesssim 0.4$. Panel a) shows a snapshot where the dense (left) region is colder, a strong non-equilibrium signature. The interface roughness is quantified in panel b) by the static height correlation $S_h(k)\sim k^{-1.5}$, yielding $\chi\simeq0.25$. This roughness is significantly reduced compared to the equilibrium value $\chi^{\rm eq} = 1/2$. Panel c) presents the time evolution of the interface width squared for various system sizes. During coarsening, we observe a slow growth of the interface width squared $W^2(t)\sim t^{0.18}$, incompatible with the equilibrium scaling $\sim t^{1/3}$. Combining this with $\chi\simeq 0.25$ yields a dynamical exponent $z \simeq 2\chi/0.18 \simeq 2.78$ via Eqs.~\eqref{eq: observable}. These exponents are confirmed in panels d) and e), which show measurement of $\tau_{ss}$ and $W_{ss}^2\equiv W^2(t\to\infty)$ as a function of $L$. Within finite-size limitations, our measured exponents are therefore consistent with the $|\bm q|$KPZ scaling.

\textit{(ii) wet-$|\bm q|$KPZ scaling}. In the middle row ($\Gamma = 0$), the momentum is locally conserved due to the absence of a Langevin bath. This additional conservation law is expected to place the system in the wet-$|\bm q|$KPZ universality class, characterized by $z^{\rm wet}=1$ and $\chi^{\rm wet}=1$. As shown in panel f), the interface is visibly rougher than in the previous case. This is quantified in panels g) and j), where $S_h(k)\sim k^{-3}$ and $W_{ss}^2 \sim L^2$ are found, both yielding $\chi \simeq 1$, consistent with theoretical expectations. The dynamics in panel h) and i), however, show deviation from standard coarsening behavior: the interface width does not exhibit clear power-law growth. This is consistent with direct simulation of the wet-$|\bm q|$KPZ field equation, which reported an anomalous roughening~\cite{caballero2024interface}. Our simulations are also plagued by persistent oscillations (see Supplemental Material~\cite{supp}). For these reasons, we do not try to investigate the typical coarsening time of this system.

\textit{(iii) Hyperuniform scaling}. In the bottom row ($T=0$ and $\Gamma\neq0$), momentum is damped but still conserved by the collisions---the only source of activity. As shown in panel k), this leads to a very flat interface~\cite{maire2025hyperuniform}. This is quantified in panel l), where the static height correlation scales as $S_h(k) \sim k^{-1}$, consistent with the theoretical prediction $\chi^{\rm HU} = 0$ in $d=1$. $\chi = 0$ implies that the interface width grows only logarithmically during coarsening (panel m), preventing a meaningful extraction of $z$ from early-time dynamics. Instead, we determine $z$ from the relaxation time (panel n), obtaining $z\simeq 3$, in agreement with the theoretical prediction $z^{\rm HU} = 3$. Finally, panel o) further confirms the vanishing roughness exponent, $\chi = 0$.

\begin{figure}
    \centering
    \includegraphics[width=0.99\linewidth]{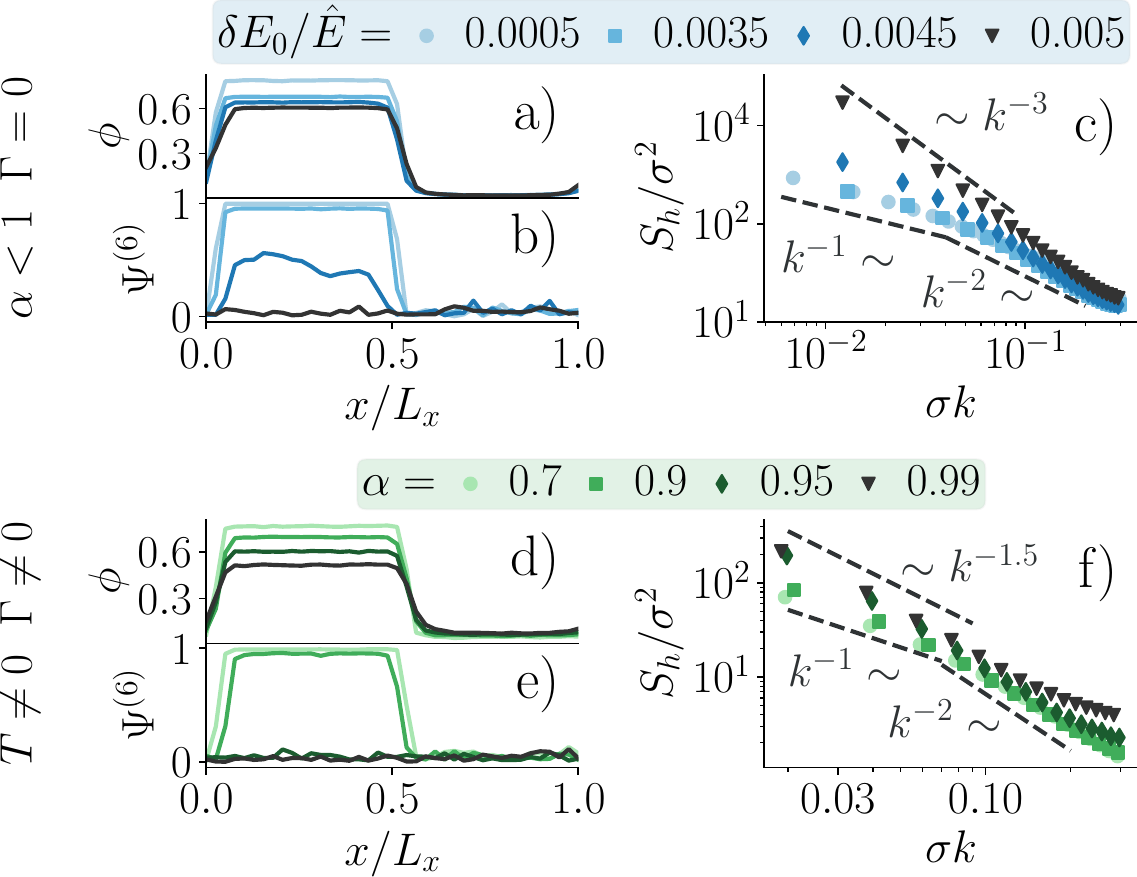}
    \caption{Simulations of a liquid-solid interface in a system with momentum conservation (top) and without momentum conservation (bottom). a), d) and b), e) are the local packing fraction and bond orientational order profiles normal to the interface, respectively. c) and f) are the static height correlations with respect to the wavevector, showing a scaling change when the dense phase undergoes a structural change. The parameters, except the ones explicitly varied, are the same as in Fig.~\ref{fig: recap}. For the system with $\Gamma = 0$, $N\simeq 2\times 10^5$, for the other $N\simeq 8\times10^4$.}
    \label{fig: liquid solid}
\end{figure}

\emph{Liquid-solid interfaces} — Having identified three universality classes for non-equilibrium liquid–gas interfaces, we now turn to non-equilibrium solid–liquid interfaces which often arise in active matter~\cite{Omar2021Phase, evans2024theory, caprini2024dynamical, bialke2015active, digregorio2018full}. While equilibrium 1D solid–liquid interfaces exhibit capillary-wave-like behavior~\cite{safran2018statistical}, out of equilibrium, the slow relaxation of the bulk may modify the interface’s static exponents. 

A hexagonal ordering naturally emerges when the dense phase exceeds a packing fraction $\phi \gtrsim 0.7$, yielding either a hexatic or a solid phase. Such a high packing fraction can be induced by tuning parameters ($\alpha \to 0$ or $\delta E_0 \to 0$) to make the dense phase effectively colder, promoting a tightly packed phase. Fig.~\ref{fig: liquid solid} presents simulations of liquid-solid interfaces, focusing on their static properties (coarsening dynamics and finite size analysis are detailed in the Supplemental Material~\cite{supp}).  The top row corresponds to momentum-conserving systems at various $\delta E_0$, where wet-$|\bm q|$KPZ scaling is expected for liquid-gas interfaces. Panel a) shows the local packing fraction profile in the direction normal to the interface. As expected, lowering $\delta E_0$ increases the dense phase packing. In panel b), we quantify the hexagonal-like ordering using the bond-orientational order parameter, defined for each particle $i$ as $\psi_i^{(6)}=N_i^{-1}\sum_{j} e^{i6\theta_{ij}}$, where $N_i$ is the number of neighbors of $i$ and $\theta_{ij}$ is the angle between particle $i$, its neighbors $j$ (with $|\bm r_i - \bm r_j| < 1.3\sigma$), and an arbitrary reference axis. At low $\delta E_0$, $\Psi^{(6)} =N^{-1}|\sum_i \psi_i^{(6)}| \to 1$, indicating a well-ordered hexagonal packing on these scales. At higher $\delta E_0$, the system remains disordered, and $\Psi^{(6)}$ remains small. For the intermediate case ($\delta E_0/\hat E = 0.0045$), the average bond-orientational parameter is reduced because contributions from slowly coarsening multiple hexatic domains partially cancel one another. In panel c), we show the interface fluctuations for these various systems. As expected, the system with a liquid dense phase follows the wet-$|\bm q|$KPZ scaling $\chi\simeq 1$, but as the dense phase orders, the spectrum shifts toward $\chi \simeq 0$. In the bottom row of Fig.~\ref{fig: liquid solid}, we perform the same analysis but for a system without momentum conservation. As shown in panel f), when the system orders (for $\alpha \leq 0.9$, as seen in panels d) and e)), the static height correlations deviate from the expected $|\bm q|$KPZ scaling and $\chi \to 0$. This qualitatively mirrors the behavior observed in the momentum-conserving system. The system with center-of-mass conservation is also found to remain at $\chi = 0$. Thus, the change of structure in the dense phase coincides with a universal shift of $\chi$ from its liquid–gas value to $0$.

\begin{figure}
    \centering
    \includegraphics[width=0.99\linewidth]{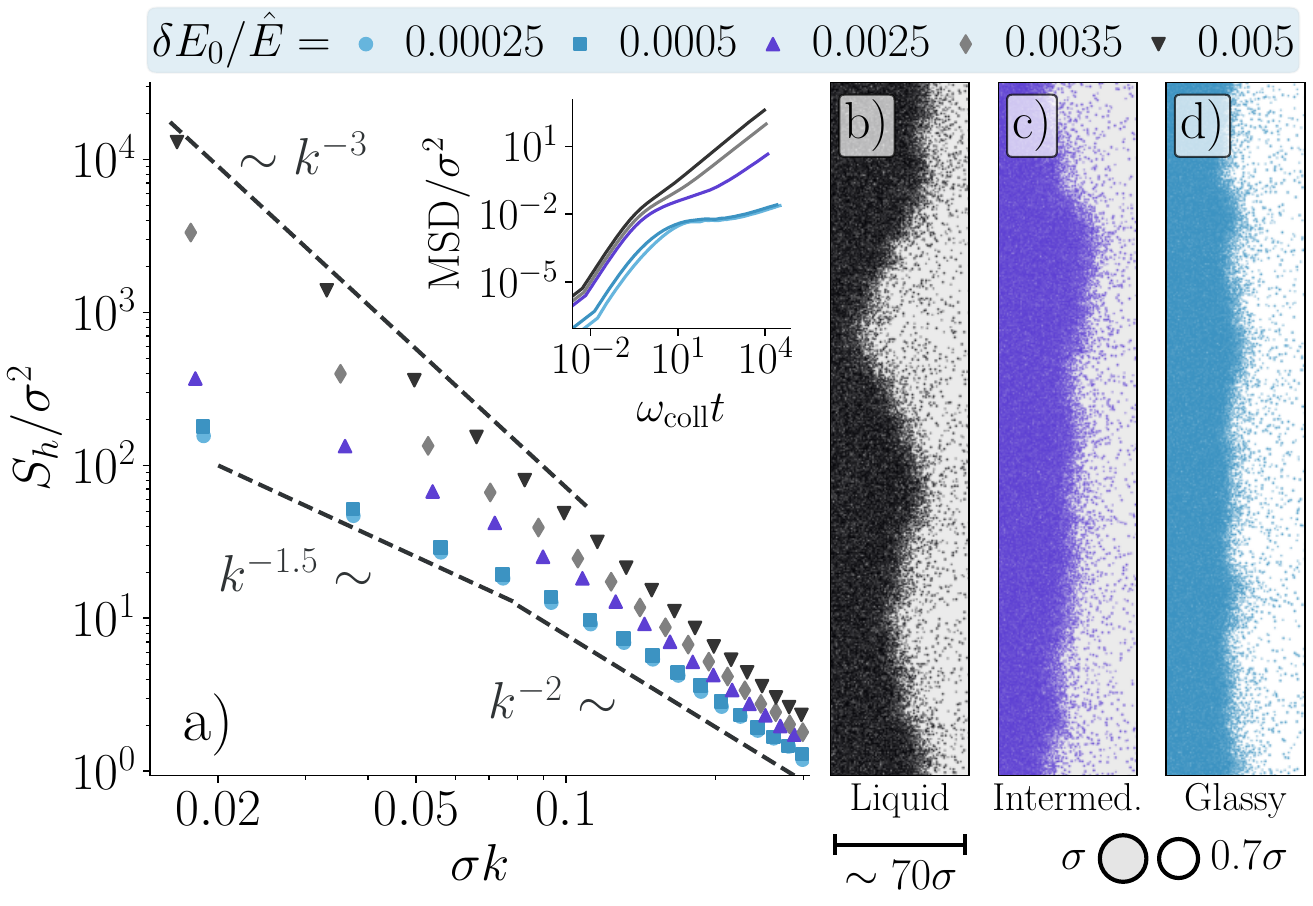}
    \caption{Simulations of a liquid–glass interface in a binary hard-disk mixture with momentum conservation. a) Static height correlations versus wavevector show a change in scaling as the dense phase becomes glassy, as measured in the inset showing the mean-square displacement (MSD) of homogeneous systems at the corresponding dense-phase densities, plotted against time normalized by the collision rate $\omega_{\rm coll}$. b–d) Representative interface profiles for low, intermediate, and high $\delta E_0$. All parameters match those in Fig.~\ref{fig: recap}, except for the varied $\delta E_0$ and $\alpha = 0.9$. $2.5\times10^5\leq N\leq 3.5\times10^5$. The mixture is equimolar with a size ratio of 0.7, and $\sigma$ refers to the large particle diameter.}
    \label{fig: liquid glass}
\end{figure}

\emph{Liquid-glass interfaces} — We now analyze liquid-glass interfaces, which are highly relevant in biological systems~\cite{kang2021novel, parry2014bacterial, mongera2018fluid, sadhukhan2024perspective}. Similar to solids, the dynamical arrest of glasses and their low-energy excitations~\cite{ goldbart2025statisticalfieldtheoryequilibrium, gartner2016nonlinear, hu2022origin} may influence the static scaling of non-equilibrium interfaces. To suppress crystalline-like structures and enable liquid–glass coexistence, we replace the monodisperse system with an equimolar binary mixture of large and small hard particles of size ratio 0.7, both governed by the same equations of motion~\cite{supp}. Denser phases are accessed by reducing the minimum injected energy $2\delta E_0$. Simulations for a momentum-conserving system are presented in Fig.~\ref{fig: liquid glass}a). At high $\delta E_0$, the scaling exponent $\chi^{\rm wet} = 1$ is recovered, as expected for an interface separating liquid and gas phases. However, as $\delta E_0$ is reduced, the dense phase undergoes dynamical arrest, evidenced by the plateau in the mean square displacement (inset). This coincides with a decrease of $\chi$ from $1$ to $0.25$. A more stable glass might yield a smaller $\chi$, potentially closer to the value found for the liquid-solid interface. As expected, the resulting liquid-glass interface is visually less rough than the liquid-gas interface, with typical configurations shown in Fig.~\ref{fig: liquid glass}b–d). We thus find a phenomenology consistent with that observed for liquid-solid interfaces: the interface roughness decreases when slow dynamics develop in the bulk of the dense phase. For both models, diffusion may be restored in the bulk of the dense phase on large timescales, potentially recovering standard liquid–gas scaling at very small $\bm k$.

\emph{Conclusion} - We introduced a minimal hard-disk model driven by active collisions that realizes three distinct non-equilibrium liquid-gas universality classes of interfacial fluctuations, each tied to specific bulk conservation laws. We also uncovered that slow dynamics arising from solid-like ordering or glassiness in the denser phase can modify the interface properties. This model directly maps onto quasi-2D vibrated granular systems undergoing phase separation, offering an experimentally accessible platform for probing non-equilibrium interfacial physics~\cite{maire2025dynamical}. Unlike many active matter systems---where long persistence lengths and interfacial polarization might necessitate a large system to observe universal scaling~\cite{Pérez-Bastías_Soto_2025, caprini2020spontaneous, Lee_2013, hermann2020active}---this granular system may offer a more accessible path to asymptotic behavior. 

To go beyond our studies, it would be valuable to develop a theory that predicts the new measured interfacial exponents and that would relate them to the slow relaxation of the bulk. Notably, it would be interesting to determine how the large-scale scaling of the interface is affected when bulk diffusion is restored at long length scales, either by defects in the solid or by the weak stability of the glass.
Building on the 1D flat interfaces studied here, exploring two-dimensional or curved interfaces---relevant for biological systems~\cite{kuroda2025effects, hyde1996language, schamberger2023curvature, sabari2020biomolecular, brangwynne2009germline, molliex2015phase, liu2025nonequilibriumdynamicsmembranelessactive, zwicker2025physics}, finite-size effects~\cite{luu2013capillarylike}, and nucleation~\cite{cates2023Nucleation, langford2024mechanics, zakine2024unveiling, richard2016nucleation, sarma2025dynamic}---offers an interesting direction. For example, nucleation driven by active collisions may exhibit strong non-equilibrium behavior~\cite{maireThesis}, distinct from that in conventional active matter~\cite{cates2023Nucleation}. More generally, interfaces with additional internal order (nematic, smectic, or polar) could reveal new universality classes, offering an exciting avenue for future study.

\section{Acknowledgements}

We gratefully acknowledge insightful discussions with Ludovic Berthier and Leonardo Galliano, as well as valuable exchanges with Adhvik Jaghannatan, Lila Sarfati, Julien Tailleur, Cesare Nardini, and Ananyo Maitra.

\bibliography{bib}

\clearpage
\onecolumngrid
\appendix

\begin{center}
\Large{\textbf{Supplemental Material}}
\end{center}

\section{Simulation methods}

\begin{figure}[!ht]
    \centering
    \includegraphics[width=0.9\linewidth]{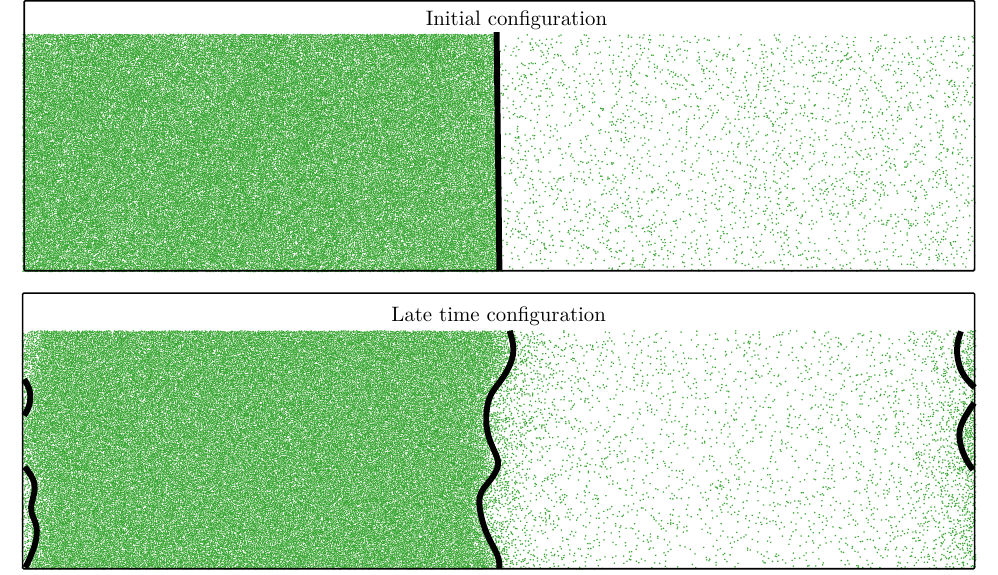}
    \caption{Initial and late time configuration of a typical system.}
    \label{fig: cartoon}
\end{figure}

We numerically integrate Eqs.~\eqref{eq: time evol} and \eqref{eq: energy change at collision} using an event-driven molecular dynamics algorithm~\cite{smallenburg2022efficient}, modified to include a Langevin bath via discrete-time kicks.

For simplicity, we first consider the case $T = 0$. In this regime, two particles $i$ and $j$ collide after a time:
\begin{equation}
    \delta\tau^{\rm coll}_{ij} = -\log\left(1-\Gamma\delta \tau_{ij}\right)/\Gamma\quad\text{with}\quad \delta \tau_{ij}=\frac{-b - \sqrt{b^2 - \bm v_{ij}^2(\bm r_{ij}^2-(\sigma_i+\sigma_j)^2/4)}}{\bm v_{ij}^2},
    \label{eq: coll time}
\end{equation}
where $b = \bm r_{ij}\cdot \bm v_{ij}$ and $\bm r_{ij}=\mathbf r_i-\mathbf r_j$ and $\bm v_{ij}=\bm v_i-\bm v_j$ are respectively the relative position and velocity of particles $i$ and $j$ at the moment the subsequent collision time is computed. $\sigma_i$ is the diameter of particle $i$ and $\delta \tau_{ij}$ is the collision time between two particles when $\Gamma\to 0$.

At each collision, particle velocities are updated according to the energy change given in Eq.~\eqref{eq: energy change at collision}, leading to the post-collision velocities $\bm v'$:
\begin{equation}
    \bm v_i'= \bm v_i - \dfrac{\bm v_{ij}\cdot \hat{\bm\sigma}_{ij}-\sqrt{\alpha^2(\bm v_{ij}\cdot \hat{\bm\sigma}_{ij})^2+4\Delta E^+/m}}{2}\hat{\bm\sigma}_{ij}
    \quad \text{and}\quad
    \bm v_j'= \bm v_j + \dfrac{\bm v_{ij}\cdot \hat{\bm\sigma}_{ij}-\sqrt{\alpha^2(\bm v_{ij}\cdot \hat{\bm\sigma}_{ij})^2+4\Delta E^+/m}}{2}\hat{\bm\sigma}_{ij},
    \label{eq: collision rule}
\end{equation}
with $\hat{\bm\sigma}_{ij}=(\bm r_i - \bm r_j)/|\bm r_i - \bm r_j|$. A collision occurs only when particles are approaching one another, i.e., when $\bm v_{ij}\cdot \hat{\bm\sigma}_{ij}<0$. Moreover, provided that  $\Delta E^+\geq0$ and $\alpha>0$, the collision rule guarantees that $\bm v_{ij}'\cdot \hat{\bm\sigma}_{ij}>0$. Therefore, after the collision, the particles move apart along the line of centers, preventing any overlap.

For a binary mixture, the same collision rule holds except that they collide at $|\bm r_i - \bm r_j|=(\sigma_{i}+\sigma_j)/2$, as indicated in Eq.~\eqref{eq: coll time}.

For $T \neq 0$, each particle receives stochastic kicks every $\delta t_{\rm noise}$, drawn from a Gaussian distribution of variance $2\Gamma T \delta t_{\rm noise}/m$~\cite{ma2017fluctuation}. To properly resolve the dynamics, we require $\delta t_{\rm noise} \ll \tau^{\rm coll}$, where $\tau^{\rm coll}$ is the mean free flight time between successive collisions.

For each parameter set, we first performed preliminary simulations in an elongated box to determine the coexistence densities. We initialized two half-box slabs at trial packing fractions, yielding an initially flat interface with a density field
\begin{equation}
    \phi_{t=0}(x,y)\simeq \big(\phi_{\rm dense}^{\rm coexistence}-\phi_{\rm dilute}^{\rm coexistence}\big)\Theta(x-L_x/2)+\phi_{\rm dilute}^{\rm coexistence},
\end{equation}
where $\Theta$ is the Heaviside function. Both phases were seeded from independent amorphous configurations prepared by starting from a dilute system, and growing the particles until the target density is reached~\cite{smallenburg2022efficient}. We then evolved the combined system until the bulk densities on each side reached stationary values. These densities (and the corresponding phases) were recorded, and we ran a short additional simulation initialized at the measured coexistence values to verify that the interface remained close to $x=L_x/2$, i.e., that the system lies near the center of the coexistence region. If not, we updated the trial densities and repeated until convergence, which fixes the total packing fraction $\phi=\pi\sigma^2N/(4L_xL_y)$. Due to strong finite-size effects, the coexistence densities depend on system size and must be tuned for each $L_x, L_y$. During these runs, we also monitored whether the dense phase underwent a phase change: when large hexagonal domains formed, we classified the phase as solid-like.

Final production runs were then started by initializing, at $t=0$, a flat interface using the converged homogeneous coexistence densities. The phases were initialized in their previously identified states (solid-like or liquid), except for $\delta E_0/\hat E=0.0045$ in Fig.~\ref{fig: liquid solid}, where the dense phase was initialized as a liquid and slowly developed hexagonal patches for $t>0$. We keep this as an instructive example, showing that a system with multiple hexatic domains behaves similarly to one with a single well-defined orientation.

When a solid-like initialization is required, we start from a perfect hexagonal lattice, which relaxes either to a solid with long-range orientational order or to a hexatic phase with quasi-long-range order, both with defects. While the steady state is insensitive to this initial relaxation, early-time coarsening may be affected.

The one-dimensional interface profile $h(y)$ is obtained by extracting the local packing fraction field $\phi(x, y)$ through binning via squares of length $5\sigma$. The position of the interface $h(y)$ is placed by fitting each $y$-slice to a hyperbolic tangent: $\phi(x, y)=A\tanh[(x-h(y))/B] + C$. The interface coarsens with time and reaches a stationary regime at a time $\tau_{ss}$, which is numerically obtained by looking at the intersection time between the early time power law of $W^2(L, t)$ and its stationary value $W^2(L, t\to\infty)$.

The initial and late time states of a typical system are illustrated in Fig.~\ref{fig: cartoon}.

\section{Large finite size effect with \texorpdfstring{$\Gamma = 0$}{Gamma=0} and density oscillations}

\begin{figure}[!ht]
    \centering
    \includegraphics[width=0.9\linewidth]{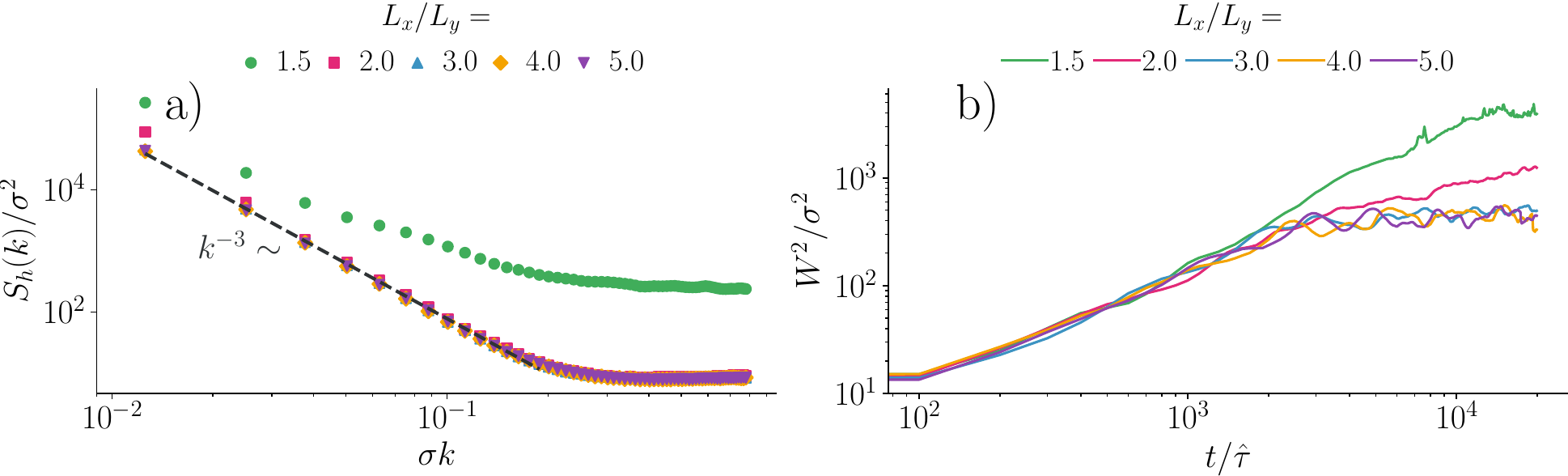}
    \caption{Quantification of the finite size effects for the system with $\Gamma= 0$ and $L_y=500\sigma$. a) Static height correlation function for various system sizes. b) Coarsening of these different system sizes with time. The parameters are the same as those used in Fig.~\ref{fig: recap} of the main text for the system with $\Gamma=0$.}
    \label{fig: finite size}
\end{figure}

To isolate bulk-driven interfacial fluctuations, the simulation box must be sufficiently long in the direction perpendicular to the interface~\cite{grant1983fluctuating}, ensuring that the two interfaces do not interact. For systems with $\Gamma = 0$, finite-size effects become particularly severe. As shown in Fig.~\ref{fig: finite size}a) where $L_x$ is varied for a fixed $L\equiv L_y$. $S_h$ varies significantly for $L_x < 3L_y$. Correspondingly, Fig.~\ref{fig: finite size}b) shows anomalous coarsening dynamics in such undersized systems, with the interface roughness growing faster than in the largest system sizes. This issue worsens with increasing $L_y$ as the critical aspect ratio $L_x/L_y$ required to suppress finite-size artifacts also increases. Thus, larger interfaces demand even more elongated simulation boxes to reach the scaling regime. This result is perhaps not unexpected, as $\chi = 1$ marks the stability limit for an interface in the thermodynamic limit. Indeed, given the scaling $\sqrt{W^2_{ss}(L_y)} \sim L_y^{\chi}$, $\chi = 1$ implies the interface width increases proportionally to its base length. Consequently, for $\chi = 1 + \varepsilon$ (with $\varepsilon > 0$), the interface becomes unstable for a given $L_y=L'_y$ such that $\sqrt{W^2(L_y')}>L_x$, if the aspect ratio $L_x/L_y$ is held fixed.

\begin{figure}[!ht]
    \centering
    \includegraphics[width=0.99\linewidth]{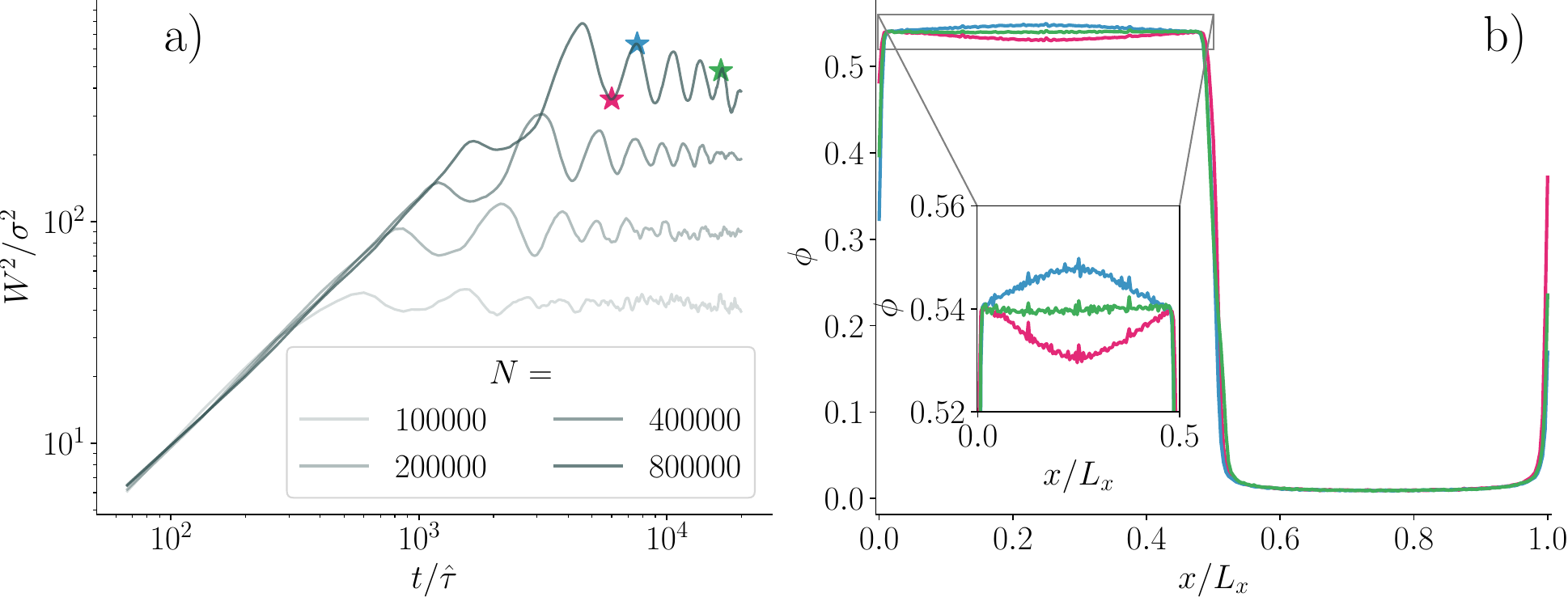}
    \caption{Spurious oscillations due to initial force imbalance at the boundaries. a) Coarsening with respect to time for different system sizes. b) Density profile along the direction perpendicular to the interface at different times corresponding to the stars marked in a). The parameters are the same as those used in Fig.~1 of the main text for the system with $\Gamma=0$.}
    \label{fig: compressible waves}
\end{figure}

However, increasing $L_x$ introduces two complications: the simulation time strongly increases and standing waves emerge. Indeed, Fig.~\ref{fig: compressible waves}a) reveals oscillations in the mean interface width, which become increasingly prominent with system size. These oscillations correspond to standing compressive waves in the density field, as shown in Fig.~\ref{fig: compressible waves}b.) They originate from the mismatch between our initial condition---a sharp flat interface---and the smooth mechanical equilibrium profile expected due to surface tension, which resembles a $\tanh(x/\zeta)$ interpolation between bulk phases. The resulting force imbalance at the phase boundaries launches compressive waves that decay only weakly in the absence of external damping ($\Gamma=0$).

This effect appears negligible for interfaces involving solid or glassy phases. Nevertheless, for gas-liquid systems, these density waves dominate interfacial coarsening, masking the intrinsic scaling behavior at early time and making it impossible to extract $z$. The examples provided here are exaggerated by purposely initializing the density field with a relatively large density mismatch compared to the measured coexistence densities.

\section{Coarsening of liquid-solid interfaces}

Coarsening dynamics in systems undergoing liquid-solid phase separation are significantly more challenging to characterize than in liquid-gas systems. An accurate initialization is important: the solid phase should ideally be unstrained, for instance, using a direct-coexistence protocol~\cite{smallenburg2024simple}. In this work, we did not implement such equilibration procedures, which may affect quantitative details of the observed coarsening. Nonetheless, we report here the qualitative trends across the different systems studied in the main text.

We first consider how coarsening changes as parameters are varied to induce a structural change in  the dense phase.

\begin{figure}[!ht]
    \centering
    \includegraphics[width=0.99\linewidth]{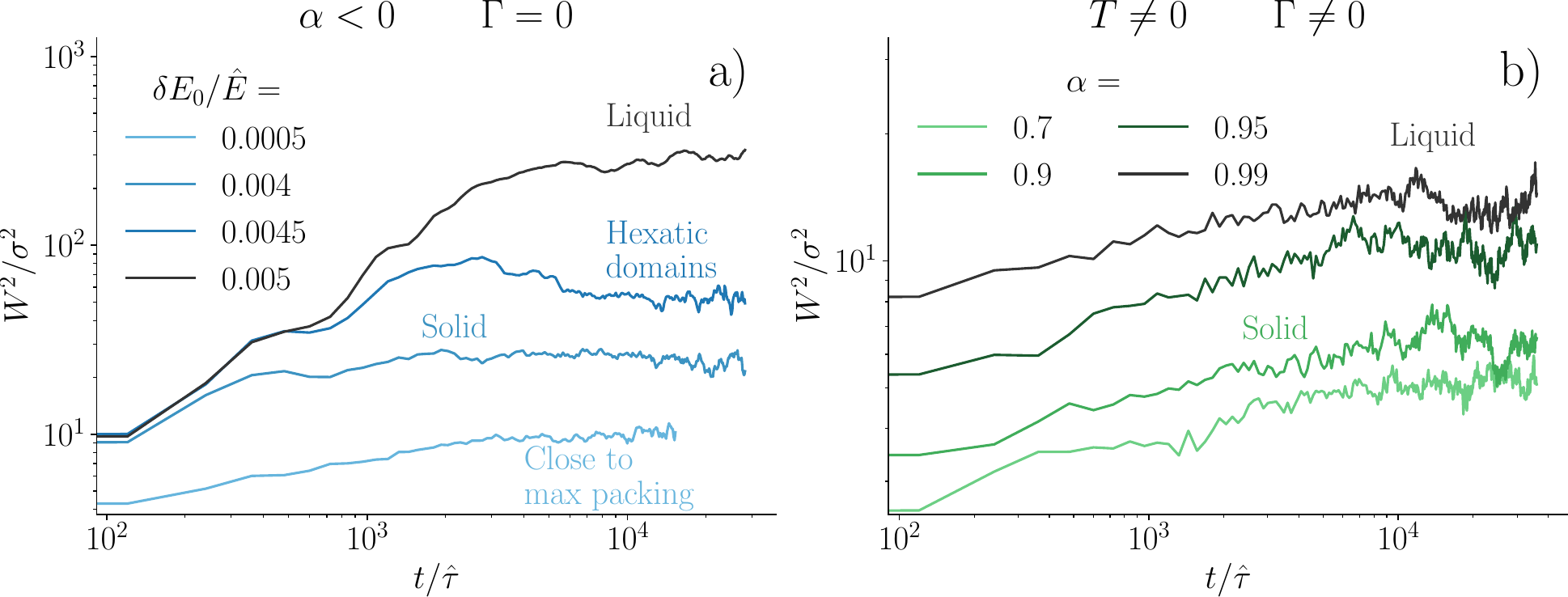}
    \caption{Variation of the coarsening when the system changes from a liquid-gas to a liquid-solid phase separation. Left is a system with momentum conservation, right is without momentum conservation. The parameters are the same as those used in Fig.~\ref{fig: liquid solid}.}
    \label{fig: coarsening liquid solid 1}
\end{figure}

The most pronounced changes are observed for $\Gamma = 0$, shown in Fig.~\ref{fig: coarsening liquid solid 1}a). For large $\delta E_0$, the system remains in a liquid-gas regime and exhibits coarsening behavior similar to that discussed in the main text (note that we did not take any precaution to partially eliminate the spurious standing waves). As $\delta E_0$ is decreased, the dense phase begins to undergo a phase change. Intermediate values produce systems with multiple hexatic domains that coarsen over time, as the initial state was roughly a disordered fluid at the same density. Interestingly, $W^2(t)$ becomes non-monotonic: the early-time increase corresponds to a predominantly liquid-gas interface, while the subsequent decrease reflects the formation of large hexatic domains that suppress interfacial fluctuations. We note that the $W^2$ becomes roughly constant at large time while the hexatic domains are still present and coarsening. We conclude that they coarsen slowly compared to the interface fluctuations. Further reduction of $\delta E_0$ yields a fully solid dense phase with markedly slower coarsening. For these systems, the dense phase was initialized with an ideal hexagonal solid. 

In Fig.~\ref{fig: coarsening liquid solid 1}b), we provide the same figure but for the system following $|\bm q|$KPZ for the liquid-gas. Once again, ordering leads to a significantly reduced interface width. However, the coarsening phase at early time remains similar.

\begin{figure}[!ht]
    \centering
    \includegraphics[width=0.99\linewidth]{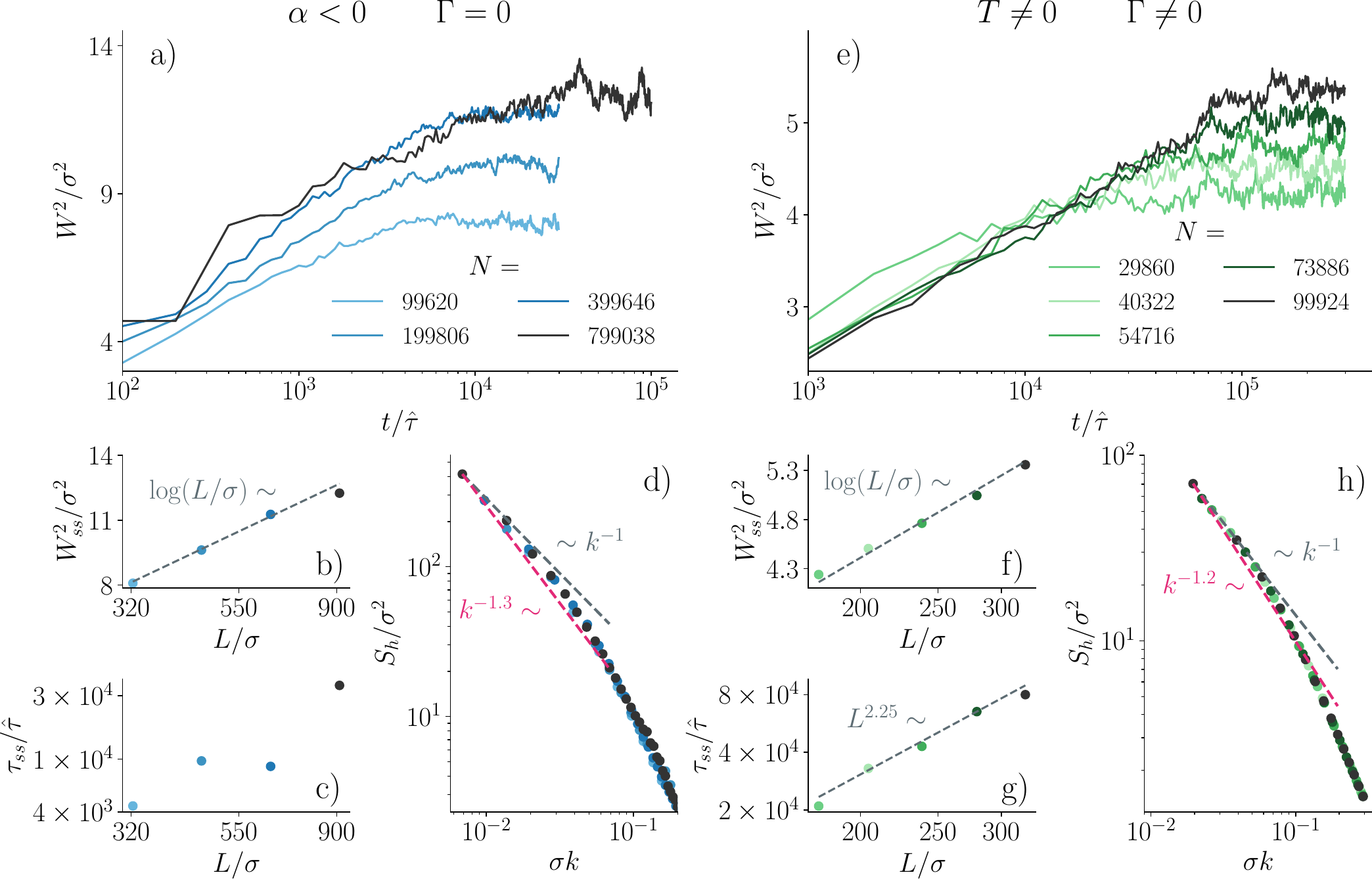}
    \caption{Finite size analysis for the systems with a liquid-solid phase separation. The left panels correspond to systems with momentum conservation, while the right panels correspond to systems without momentum conservation. a) and e) are the evolution of the mean width squared as a function of time. b) and f) are the stationary values of $W^2$ as a function of $L$. c) and g) are the time required to reach this steady state value. d) and h) are the static height correlation in the steady state.  For both systems, the parameters are the same as those used in Fig.~\ref{fig: liquid solid}, which lead to the densest glassy phase.}
    \label{fig: coarsening liquid solid 2}
\end{figure}

We now perform a finite-size analysis of systems with the smallest values of the control parameter, corresponding to the most densely packed solids.
We first consider the system with $\Gamma = 0$ (left-hand side of Fig.~\ref{fig: coarsening liquid solid 2}). As shown in Fig.~\ref{fig: coarsening liquid solid 2}a) and c), the interface coarsens in a manner that prohibits a clear extraction of $z$. The steady state $W^2$ can, however, be extracted and is given in panel b). The approximate logarithmic scaling suggests $\chi \simeq 0$. This trend is consistent with the height structure factor $S_h(k)$ shown in panel d), which predicts $0\lesssim\chi\lesssim0.15$.

In contrast, the system with $\Gamma \neq 0$ (right-hand side of the figure) shows more regular behavior. As seen in panel e), the coarsening appears well-behaved, and analysis of panels f) and g) yields estimates of $0\lesssim\chi \lesssim 0.1$ and $z \simeq 2.25$. 

\end{document}